\documentclass[%
 article,two column,preprintnumbers,
superscriptaddress,
amsmath,amssymb,aps,
]{revtex4-2}
\usepackage{graphicx}
\usepackage{dcolumn}
\usepackage{bm}
\usepackage{bbold}
\usepackage{calc}
\usepackage{changes}
\usepackage{xcolor}
\usepackage{hyperref}
\usepackage{color}
\usepackage{xcolor,colortbl}
\usepackage{comment}
\newcommand {\nc} {\newcommand}

\newcommand{\al}{\alpha}

\newcommand{\dt}{\delta}

\newcommand{\mpi}{m_{\pi}}

\newcommand{\dslash}[1]{#1 \llap{/\kern-0.5pt}}
\newcommand{\Dslash}[1]{#1 \llap{/\kern+1.5pt}}
\newcommand{\DDslash}[1]{#1 \llap{/\kern+2.3pt}}
\newcommand{\dslashh}[1]{#1 \llap{/\kern+1pt}}

\newcommand{\beq}{\begin{equation}}
\newcommand{\eeq}{\end{equation}}

\newcommand{\bea}{\begin{eqnarray}}
\newcommand{\eea}{\end{eqnarray}}
\newcommand{\bma}{\begin{pmatrix}}
\newcommand{\ema}{\end{pmatrix}}
\newcommand{\nn}{\nonumber}

\newcommand{\nocontentsline}[3]{}
\newcommand{\tocless}[2]{\bgroup\let\addcontentsline=\nocontentsline#1{#2}\egroup}

\newcommand{\nsat}{n_{\rm sat}}

\newcommand{\XXint}[3]{{\setbox0=\hbox{$#1{#2#3}{\int}$}
\vcenter{\hbox{$#2#3$ }}\kern-.65\wd0}}

\newcommand{\remark}[1]{}

\definecolor{darkgreen}{rgb}{0,0.5,0}
\definecolor{darkblue}{rgb}{0,0,0.5}
\definecolor{darkred}{rgb}{0.5,0,0}
\definecolor{beige}{rgb}{0.7,0.4,0.3}

\graphicspath{ {./Figures/} }

\newcommand{\RNum}[1]{\uppercase\expandafter{\romannumeral #1\relax}}

\nc {\IR} [1]{\textcolor{red}{#1}}
\nc {\IB} [1]{\textcolor{blue}{#1}}
\nc {\IM} [1]{\textcolor{magenta}{#1}}
\hypersetup{colorlinks=true,linkcolor=blue,citecolor=blue,filecolor=magenta,urlcolor=blue,pdftitle={Overleaf Example},pdfpagemode=FullScreen}

\begin{document}

\preprint{INT-PUB-24-059}

\title{A New Class of Three Nucleon Forces and their Implications}
\author{Vincenzo Cirigliano}%
\email{cirigv@uw.edu}
\affiliation{
Institute for Nuclear Theory, University of Washington, Seattle, WA 98195, USA
}
\author{Maria Dawid}%
\email{mdawid@uw.edu}
\affiliation{
Institute for Nuclear Theory, University of Washington, Seattle, WA 98195, USA
}
\author{Wouter Dekens}%
\email{wdekens@uw.edu}
\affiliation{
Institute for Nuclear Theory, University of Washington, Seattle, WA 98195, USA
}
\author{Sanjay Reddy}
\email{sareddy@uw.edu}
\affiliation{
Institute for Nuclear Theory, University of Washington, Seattle, WA 98195, USA
}%

\date{\today}

\begin{abstract}
We identify a new class of three-nucleon forces that arises in the low-energy effective theory of nuclear interactions including pions.  We estimate their contribution to the energy of neutron and nuclear matter and find that it can be as important as the leading-order three-nucleon forces previously considered in the literature. The magnitude of this force is set by the strength of the coupling of pions to two nucleons and is presently not well constrained by experiments. The implications for nuclei, nuclear matter, and the equation of state of neutron matter 
are briefly discussed.        
\end{abstract}

\maketitle
\section{Introduction}
The development and application of effective field theory (EFT) methods to low-energy nuclear processes has had a profound impact on our understanding of nuclei and dense nuclear matter encountered in neutron stars \cite{Epelbaum:2008ga,Hammer:2019poc}. Pioneering work by Weinberg laid the foundations for describing interactions between nucleons in a systematic EFT approach \cite{Weinberg:1990rz,Weinberg:1991um} 
that exploits the separation  of energy scales and is consistent with the symmetries of QCD. 
Chiral EFT describes interactions between nucleons at energy scales comparable to the pion mass but much smaller than the mass of the heavier mesons such as the $\rho$ meson. This theory organizes the operators involving nucleon and pion fields in a momentum expansion in which higher-order terms are suppressed by powers of ($p/\Lambda_B$ or $m_\pi/\Lambda_B$), where $p$ is the typical momentum scale 
of the considered process,
$m_\pi$ is the pion mass, and  $\Lambda_B$ is the 
breakdown scale associated with the EFT. The numerical value of the coefficients of the operators in the expansion, called the Low Energy Constants (LECs), are fixed by matching to experiments. A particularly attractive feature of the EFT approach to nuclear interactions is that it includes many-body forces naturally and organizes them in a useful hierarchy that is consistent with the momentum expansion of the two-nucleon force (2NF). 
\looseness-1

The three-nucleon force (3NF) plays a critical role in describing observables such as the nuclear binding energies and radii, and the equation of state (EOS) of dense neutron-rich matter in neutron stars. Although the magnitude of its contribution to these quantities depends on the resolution or regularization scale of the calculation, often defined by the momentum cut-off denoted by $\Lambda$, for the typical values of  $\Lambda \simeq 450-500$ MeV, the 3NF is essential to stabilize both nuclei and neutron stars. Without 3NFs, the nuclear binding energy and saturation density would be too high, and the maximum mass and neutron star radii would be too low.

The leading 3NF appears at the next-to-next-to-leading order (N$^2$LO) in the EFT expansion~\cite{vanKolck:1994yi,Epelbaum:2002vt}. At N$^2$LO, the 3NF in Chiral EFT contains a long-range contribution in which two pions are exchanged between nucleons, an intermediate-range interaction, and a short-range contribution. The LECs associated with intermediate-range interaction, and short-range contributions, called $c_D$ and $c_E$, respectively, are determined by fitting to properties of tritium and helium. Interestingly, the LECs associated with the leading long-range part of the 3NF, which is well known as the Fujita Miyazawa force \cite{Fujita_Miyazawa:1957}, can be determined by the analysis of pion-nucleon scattering in Chiral perturbation theory 
and does not rely on data from multi-nucleon systems.\looseness-1 

In this letter, we identify new loop contributions to the 3NF, which arise from operators that involve four nucleon fields and two pion fields. One of these operators is related by chiral symmetry to the quark mass-dependent four-nucleon contact operator, which can be interpreted as the nuclear $\sigma$ term \cite{Beane:2013kca}. 
Arguments based on the renormalization 
of the nucleon-nucleon spin-singlet s-wave scattering amplitude~\cite{Kaplan:1996xu}  
warrant that the associated LEC, denoted by $D_2$, is 
significantly 
larger than 
what is expected in Weinberg's power counting~\cite{Weinberg:1990rz,Weinberg:1991um}.
Two other contributions to the 3NF arise from terms involving derivatives of the pion fields, with couplings denoted by $E_2$ and $F_2$. 
These LECs are not well constrained now, but $D_2$ can in principle be determined by performing lattice QCD calculations of the nucleon-nucleon scattering length as a function of $\mpi$, while all three can be constrained by pion scattering of two nucleons in the spin singlet state, 
or the properties of light nuclei. \looseness-1

Based on the renormalization group scaling, 
we find that the new 3NFs appear at N$^3$LO,  
while they would appear at  N$^5$LO in Weinberg's counting. 
Moreover, 
due to numerical enhancements  
the new 3NFs induce effects comparable to those of the leading  N$^2$LO 3NF.
For 
values of the 
LECs 
implied by our renormalization analysis, 
the energy per particle in neutron and nuclear matter is of the order 
of $10$ MeV at nuclear saturation density and grows rapidly with increasing density. \looseness-1

\section{A quark mass dependent 3NF}
Kaplan, Savage, and Wise (KSW) in \cite{Kaplan:1996xu} showed that the 
renormalization of the leading order $NN$ amplitude  
in the $^1S_0$ channel requires introducing  quark-mass 
dependent four-nucleon interactions in the leading order Lagrangian. 
In terms of nucleons and the chiral fields, 
$N=(p,n)^T$, $\chi_+ = u^\dagger \chi u^\dagger + u \chi^\dagger u$, $\chi = 2 B_0 m_q $, and $U = u^2 = {\rm exp}(i \tau^a \pi^a/f_\pi)$ (where $f_\pi=92.4$~MeV is the pion decay constant),  
and working in the isospin limit, the required interactions can be written in chiral covariant form as follows~\cite{Mehen:1998tp},
\bea
{\cal L} &=& \left[ d_2^S \, \bar N N \, \bar N N+ d_2^T  \bar N\vec \sigma N \, \bar N \vec \sigma N\right]\,  \langle \chi_+ \rangle \,,\nn\\
&=&-\frac{1}{4}D_2 \, ( N^T P_i N)^\dagger \, ( N^T P_i N)\,  \langle \chi_+ \rangle \nn\\
&&-\frac{1}{4}D'_2 \, ( N^T P'_iN)^\dagger \, ( N^T P'_i N)\,  \langle \chi_+ \rangle ~,
\\
\langle \chi_+ \rangle &=& 
4  m_\pi^2  \left(1 - \frac{\pi^a \pi^b}{2 f_\pi^2}  \delta^{ab}  + \cdots \right) ~, 
\label{eq:d2LagPW}
\eea
where $\sigma^i\, (\tau^i)$ are the (iso)spin Pauli matrices, while $P_i=\tau_2\tau^i \sigma_2/\sqrt{8}$ and $P'_i=\sigma_2\sigma^i \tau_2/\sqrt{8}$ are the $^1S_0$ and $^3S_1$ partial-wave projection operators, respectively. 
The scalar/tensor couplings  ($d_2^S$,  $d_2^T$) and the partial-wave couplings ($D_2$ and $D'_2$)
are related by
\bea
 d_2^S = -\frac{D_2+3D'_2}{32}\,,\qquad d_2^T = \frac{D_2-D'_2}{32}\,.
\label{eq:PWtranslate}
\eea
$D_2$ and $D'_2$ could be determined from pion-nucleus scattering or lattice QCD \cite{Beane:2006mx}, but are currently poorly constrained.
In this work we estimate $D_2$ based on the scaling required by the renormalization group equation~\cite{Kaplan:1996xu} 
\beq
\frac{d}{d\ln\mu }\left[\frac{m_\pi^2 D_2}{\tilde C_0^2}\right]_{} =\frac{g_A^2 m_\pi^2 m_N^2}{64\pi^2 f_\pi^2}\,,
\label{eq:anom_dim_1}
\eeq
where $\tilde C_0$ is the leading-order contact interaction in the $^1S_0$ channel and $\mu$ is the renormalization scale in dimensional 
regularization~\footnote{Note that Ref.~\cite{Gegelia:2004pz} under-estimates  
the anomalous dimension in the RHS of Eq.~\eqref{eq:anom_dim_1} by a factor of four.  
Our direct calculation agrees with  the result of Ref.\cite{Kaplan:1996xu}}. Similar results for the logarithmic scaling with the cutoff hold in other regularization  schemes~\cite{Cirigliano:2019vdj}. 
At a generic scale the above relation implies 
\beq
D_2(\mu) \approx \frac{g_A^2 m_N^2}{64\pi^2 f_\pi^2}~\tilde{C}^2_0(\mu) \,. 
\label{eq:xi0}
\eeq
More generally, we define the dimensionless ratio 
\bea
\xi = \left| \frac{D_2}{\tilde{C}^2_0} \right|\,,
\eea
note that Eq.~\ref{eq:xi0} predicts $\xi=0.26$, and explore the range $\xi \sim 0.1 - 0.5$. 
For typical values of $\tilde C_0$, this corresponds to  $|D_2 |< 1/(5 f_\pi^4)$. 
In \cite{Beane:2002xf}, the authors introduce a related dimensionless ratio $\eta= |D_2 m_\pi^2/\tilde{C}_0|$ and suggest the range $ 1/15< \eta < 1/3$ to ensure consistency with a perturbative treatment of the pion exchange~\cite{kaplan:1998we,Kaplan:1998tg}.
Since $\xi = \eta/ (m_\pi^2 \tilde{C}_0)$, and $\tilde{C}_0 \approx 1/m_\pi^2$ in commonly employed regularization schemes, the range we explore here is roughly compatible with the range suggested in \cite{Beane:2002xf}. 
In Chiral EFT, for typical $\Lambda$ employed in calculations, $\tilde{C}_0 \approx -5$ fm$^2$, and $\xi< 0.5$ predicts $|D_2| \lesssim 10$ fm$^4$.  

When pions are treated non-perturbatively, 
the LECs associated with 
the spin-triplet 
operators 
are not required for renormalization of scattering amplitudes to 
leading order~\cite{Beane:2001bc}. For this reason, 
we assume 
$D'_2 \ll D_2$ and 
neglect the contribution of the spin-triplet operators. \looseness-1

\begin{figure}[t!]
\vspace{0.1cm}
    \centering
\includegraphics[height=2.cm]{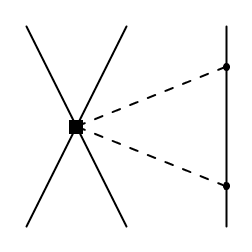} 
\hspace{0.2cm}
\includegraphics[height=2.cm]{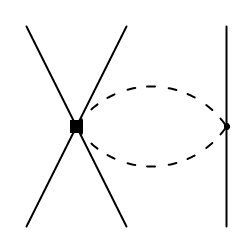} %
\hspace{0.2cm}
\includegraphics[height=2.cm]{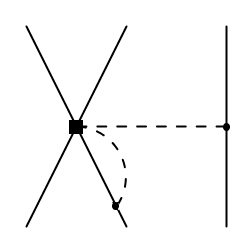} 
\caption{Diagrams that contribute to the 3NF due to the 2$\pi$-exchange interaction induced by insertions of the $D_2$ operator 
(black circle).
Solid lines represent nucleons and dashed lines represent pions.}
    \label{fig:diagram}%
\end{figure}
The $D_2$ operator induces  three-nucleon potentials through the 
Feynman diagrams shown in Fig.~\ref{fig:diagram}. In momentum space, we find 
\beq
V^{i'j'k'}_{ijk}(\vec q_1,\vec q_2,\vec q_3) = \frac{9 g_A^2  D_2 m_\pi^3 }{128\pi f_\pi^4} \kappa^{i'j'}_{ij}\dt_{kk'}~{\cal I}\left(\frac{\vec q_3^{\, 2}}{4m_\pi^2}\right)\,,
\label{eq:V3ND2}
\eeq
where $\kappa^{i'j'}_{ij}=
(d_2^S\dt_{ii'}\dt_{jj'}+d_2^T\vec \sigma_{i'i}\cdot \vec \sigma_{j'j})/(d^S_2-3d^T_2)$, and
\beq
 {\cal I}(b)= \frac{2}{3}  \left( 1 + \left(\frac{1}{2\sqrt{b}}+\sqrt{b} \right)  \tan^{-1} (\sqrt{b})\right)\,.
\eeq
Here $q_i = p_i'-p_i$ is the exchanged momentum of nucleon $i$, and $\sum q_i=0$.
Performing the Fourier transform in dimensional regularization (DR) or regulating the integrals with a small imaginary part of $r$, 
the coordinate-space potential becomes
\bea
V_{i'j'k'}^{ijk}(\vec r_2,\vec r_3) &=& -\frac{9 g_A^2 D_2 m_\pi^3 }{128\pi f_\pi^4}\kappa^{i'j'}_{ij}\dt_{kk'} \dt^3(\vec r_2)f(|\vec r_3|)\,,\nn\\
f(r)\Big|_{\rm DR}&=&- \frac{ e^{-2m_{\pi}r} }{6\pi m_\pi  r^4}(1+m_{\pi}r)^2\,,\nn\\
f(r)\Big|_{{\rm Im} (r)} &=&\frac{2}{3}\delta^3(r)- \frac{ e^{-2m_{\pi}r} }{6\pi m_\pi  r^4}(1+m_{\pi}r)^2\,.
\eea

The $D_2$ scaling in \eqref{eq:xi0} implies the above potentials appear at N$^3$LO, one order below the leading N$^2$LO  3NF. 
This expectation assumes a 
loop 
suppression of $\sim 1/(4\pi)^2$,  
even though possibly large numerical factors and powers of $\pi$ can be hard to predict~\footnote{For example, numerically large and/or $\pi$-enhanced contributions have been encountered in $\pi-N$ and $\pi$-deuteron scattering \cite{Gasser:2002am,Hoferichter:2009ez,Hoferichter:2009gn,Beane:2002wk,Liebig:2010ki,Baru:2010xn,Baru:2011bw,Baru:2012iv}. However, there is no general power-counting that can capture such enhancements, without explicit calculation.}.
Indeed, the 
result of Eq.\ \eqref{eq:V3ND2} is larger than the naive expectation by at least a factor of $\pi$. As we will see later, the $D_2$ contributions also have a stronger momentum dependence than the leading 3NF, giving further enhancement at large densities. \looseness-1

\section{3NFs from derivative terms}
The consistent renormalization of pion scattering off two-nucleons in the spin-singlet channel  also requires two other operators that describe the coupling of pions to two nucleons involving derivatives of the pion field \cite{Borasoy:2001gq,Borasoy:2003gf}.
Following Ref.\ \cite{Borasoy:2003gf} we write these additional $\pi^2 NN$ operator structures that are independent of the pion mass~
\footnote{In principle, an isospin-tensor operator of the form, $ (N^T P_a N)(N^T P_b N)^\dagger \langle u_\mu \tau^c\rangle \langle u^\mu  \tau^d\rangle(3\delta_{ac}\delta_{bd}-\delta_{ab}\delta_{cd})$ appears in the $^{1}S_0$ channel, in addition to the singlet term, $\sim \delta_{ab}\delta_{cd}$, in Eq.\ \eqref{eq:E2F2}. In fact, we find that this operator has the same anomalous dimension as that of Eq.\ \eqref{eq:E2F2RG}. We do not discuss this term any further here, as the contributions to the 3NF are only sensitive to the singlet operator.}
\bea\label{eq:E2F2}
{\cal L}_2 &=& 
\frac{1}{4}
\left[E_2 \langle( v\cdot u)^2\rangle +F_2 \langle u\cdot u-(v\cdot u)^2\rangle
\right]\,  \nn \\
& \times& ( N^T P_i N)^\dagger \, ( N^T P_i N) \,   \\
&+& \frac{1}{4}
\left[E'_2 \langle( v\cdot u)^2\rangle +F'_2 \langle u\cdot u-(v\cdot u)^2\rangle
\right]\,  \nn \\
& \times& ( N^T P'_i N)^\dagger \, ( N^T P'_i N) \, \nn
\eea
where $u_\mu = i(u\partial_\mu u^\dagger-u^\dagger \partial_\mu u)/2$. Expanding $u$ and retaining only the leading terms we find that 
\bea
{\cal L}_2&=&-\frac{2}{f_\pi^2}\sum_{\alpha =S,T}(\bar N \Gamma_\al N)\cdot (\bar N \Gamma_\al N)\,\\
&\times& \left[e^\al_2(\partial_0 \pi^A)(\partial_0 \pi^A) -f^\al_2(\vec\nabla \pi^A)(\vec\nabla \pi^A)
\right]
\,  +{\cal O}(\pi^4)\,\nn
\,
\eea
where $\Gamma_S = 1$ and $\Gamma_T = \vec \sigma$. The $S,T$ couplings are related to the partial-wave basis in analogy to Eq.\ \eqref{eq:PWtranslate}
\bea
 e_2^S &=& -\frac{E_2+3E'_2}{32}\,, \quad e_2^T = \frac{E_2-E'_2}{32}\,\nn\\
  f_2^S &=& -\frac{F_2+3F'_2}{32}\,, \quad f_2^T = \frac{F_2-F'_2}{32}\,.
\eea

We find the following renormalization group equations (RGEs) for the $^1S_0$ operators (see Ref.\ \cite{Borasoy:2003gf} for the relevant diagrams in the $^3S_1$ channel)
\bea\label{eq:E2F2RG}
\frac{d}{d\ln\mu}\left[\frac{X}{\tilde C_0^2}\right] = \gamma_X \left(\frac{m_N}{4\pi f_\pi}\right)^2\,,
\eea
with $X\in\{E_2,F_2 \}$, $\gamma_{E_2} = -(1+g_A^2)/3$, and $\gamma_{F_2} = -g_A^2/3$.
Since its RGE is similar to that of $D_2$, in what follows we will explore the same range for the size  of $F_2$, namely,  $|F_2| \leq 1/(5f_\pi^4)$.

The 3N potential can be calculated as in the case of the $D_2$ operator, and in momentum space, we find that   
\bea
V^{i'j'k'}_{ijk}(\vec q_1,\vec q_2,\vec q_3) &=& 
- \frac{15 g_A^2 m_\pi^3 }{16\pi f_\pi^4} \dt_{kk'} {\cal J}\left(\frac{\vec q_3^{\, 2}}{4m_\pi^2}\right) \nn \\
&\times &\Bigg(f_2^S\dt_{ii'}\dt_{jj'}+f_2^T\vec \sigma_{i'i}\cdot \vec \sigma_{j'j}\Bigg)\,,
\label{eq:V3NF2}
\eea
where
\beq
{\cal J}(b)= \frac{3}{5}  \left( (1+2b){\cal I}(b)+\frac{2}{3}\right)\,,
\eeq
$q_i = p_i'-p_i$ is the exchanged momentum of nucleon $i$, and $\sum q_i=0$. We neglect the contribution from the $E_2$ operator because it is proportional to the kinetic energy of the nucleons~\footnote{One might expect these contributions to become comparable to $D_2$ for Fermi momenta of ${\cal O}(300)$ MeV, at which point $E_{\rm kin}^2\sim m_\pi^2$. However, explicit calculation shows that the $E_2$ terms only become comparable for significantly larger Fermi momenta.}.

\section{Neutron and Nuclear Matter}
\label{sec:matter}
To assess the importance of the $D_2$ and $F_2$ terms in the 3NF we estimate their contributions to the energy per particle in uniform neutron and nuclear matter in perturbation theory. We shall assume that the Fermi Gas (FG) reference state well describes the momentum distribution of nucleons in the ground state. The interaction energy density is obtained by calculating the matrix element of the potential in Eq.\ \eqref{eq:V3ND2} in the FG ground state. Accounting for all possible contractions we find 
\bea\label{eq:Edensity}
\langle {\cal H}(0)\rangle  &=& \int_{\vec p_1,\vec p_2,\vec p_3}\theta(k_f-|\vec p_1|)\theta(k_f-|\vec p_2|)\theta(k_f-|\vec p_3|)\nn\\
&&\times\Bigg[V_{ijk}^{ijk}(0,0,0)-V_{ijk}^{ikj}(0,\vec p_{32},\vec p_{23}) \nn \\
&&+V_{ijk}^{jki}(\vec p_{21},\vec p_{32},\vec p_{13})
+V_{ijk}^{kij}(\vec p_{31},\vec p_{12},\vec p_{21})\nn\\
&&-V_{ijk}^{kji}(\vec p_{31},0,\vec p_{13})-V_{ijk}^{jik}(\vec p_{21},\vec p_{12},0)
\Bigg]\,,
\eea
In neutron matter the $i,j,k$ indices only run over the spins, while they also run over isospin in symmetric matter. The momentum integrals can be performed and we find that the energy per particle in neutron and symmetric matter due to the potential in \eqref{eq:V3ND2} are given by 
\bea
E^{D}_{\rm NM} &=& -\frac{g_A^2 m_\pi^9D_2}{8960\pi^5f_\pi^4}\Bigg[4u^5(7+3u^2)\cot^{-1}(1/u) \nn \\
&+& 2(3+7u^2)\log(1+u^2)  \nn \\
&-&u^2(6+11 u^2+23u^4)\Bigg] \,,
\label{eq:EDNM}
\\
\frac{E^{D}_{\rm SM}}{A}&=&\frac{3(D_2+D'_2)}{2 D_2}\left(E^{D}_{\rm NM}-\frac{ 9g_A^2 m_\pi^3 D_2}{1024\pi f_\pi^4}n_B^2\right)\,,
\eea
respectively. Here $u=k_F/m_\pi$, and $k_F=(3\pi^2 n_n)^{1/3}$ in neutron matter and in symmetric matter $k_F=(3\pi^2 n_B/2)^{1/3}$.  The energy per particle due to the potential in \eqref{eq:V3NF2} takes the form
\bea
\label{eq:EFNM}
E^{F}_{\rm NM}&& = -\frac{g_A^2 m_\pi^9F_2}{241920\pi^5f_\pi^4}\Big[
u^2\Big(138+630\pi u -735 u^2 \nn\\
&-& 587 u^4 + 240u^6-1260 u \cot^{-1}(u) \\
&+&12u^3(63+54 u^2+20u^4)\tan^{-1}(u)\Big) \nn\\ 
&-&6(23+99 u^2)\log (1+u^2) \Big]\nn
\,, \\
E^{F}_{\rm SM}&=&\frac{3(F_2+F'_2)}{2F_2}\left( E^{F}_{\rm NM}+\frac{ 15g_A^2 m_\pi^3 n_B^2}{1024\pi f_\pi^4 } F_2\right)\,.
\eea

\begin{figure}[t!]%
    \centering
\includegraphics[width=1\linewidth]{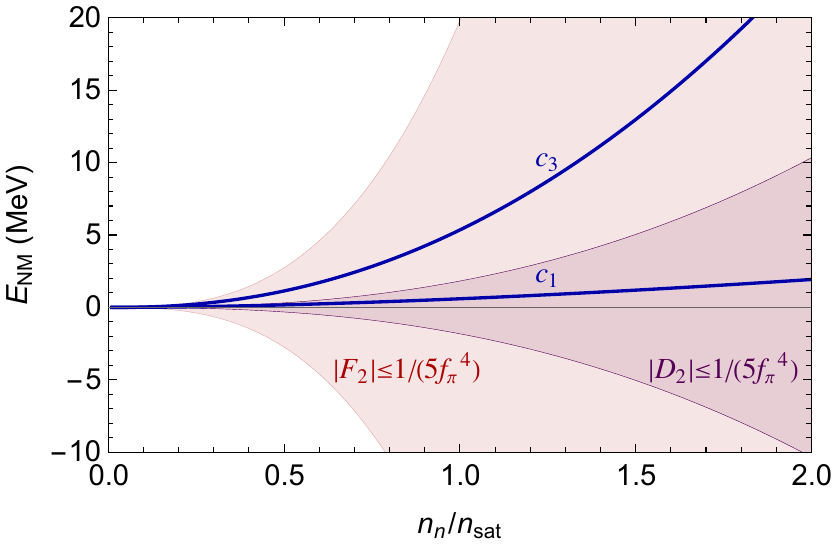} %
    \caption{ The figure shows $D_2$ and $F_2$ contributions to the energy per particle in neutron matter as a function of the density. The bands are obtained by setting $|D_2|<1/(5f_\pi^4)$ and $|F_2|<1/(5f_\pi^4)$. For reference, we show the contributions from the long-range Fujita-Miyazawa 3NF induced by the LECs $c_1$ and $C_3$,  obtained 
   by setting  $c_1=-0.81/$GeV and $c_3=-3.2/$GeV \cite{Hebeler:2009iv}.
    }%
    \label{fig:NMplot}%
\end{figure}

The resulting contributions in neutron matter are shown in Fig. \ref{fig:NMplot} for $|D_2| < 1/(5f_\pi^4)$ and $|F_2| < 1/(5f_\pi^4)$ while neglecting the $^3S_1$ couplings.
In symmetric matter, the corresponding contributions are shown in Fig.~\ref{fig:SMplot}. 
The results in these figures employ the DR result for the 3N potentials. If we instead use the regularization prescription of Refs.\ \cite{Rijken:1990qs,Reinert:2017usi}, the long-range part of the potential tends to produce smaller contributions in dense matter. For example, in the case of the $D_2$, the contributions in neutron matter differ by a factor of $E_{\rm NM}^{(\Lambda)}/E_{\rm NM}^{\rm(DR)}=\{0.10,0.29,0.59,0.78\}$ at $n=n_{\rm sat}$, for  $\Lambda=\{0.3,0.5,1,2\}$ GeV, where $\Lambda$ is the cutoff in the scheme of Refs.\ \cite{Rijken:1990qs,Reinert:2017usi}. The difference between the two schemes will be compensated by differences in LECs such as $c_E$. A fully consistent analysis will therefore have to determine $D_2$ and $F_2$, together with the usual contact terms $c_{D,E}$.\looseness-1

\section{Implications}
\label{sec:astro}
\begin{figure}[t!]%
    \centering
\includegraphics[width=1\linewidth]{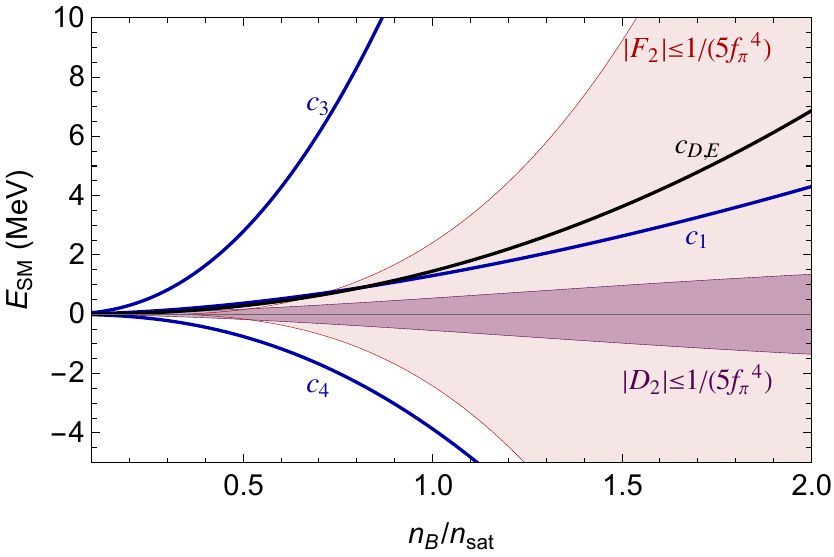} 
    \caption{
   Same as Fig.\ \ref{fig:NMplot}, but for symmetric matter.  For the single-nucleon LECs we use  $c_1=-0.76/$GeV, $c_3=-4.78/$GeV, and $c_4=3.96/$GeV, while for the $2$ and $3$ nucleon terms we 
   take $c_D = 2.08$,  and $c_E=0.23$ for $\Lambda_\chi=700$ MeV \cite{Bogner:2005sn}.}%
    \label{fig:SMplot}%
\end{figure}

For Chiral EFT to provide a useful description of nuclei and neutron-rich matter, the results in Figs.~\ref{fig:NMplot} and \ref{fig:SMplot} indicate that $D_2$ and $F_2$ will need to be determined rather accurately. In an {\it ab initio} approach, they would be deduced by fitting to the binding energies of light nuclei or from pion-nucleus scattering data. In the former approach, one would need to simultaneously determine $c_D$, $c_E$, $D_2$, and $F_2$ from light nuclei. Absent such constraints, in what follows we focus on neutron matter and adopt a phenomenological approach to correlate $D_2$ and $F_2$ using empirical information about the nuclear symmetry energy $S_0=31.7\pm 3.2$ MeV at saturation density~\cite{Li:2019xxz}. 

Since the LECs associated with the 2NF and the long-range 3NF ($c_1$ and $c_3$) between neutrons are well constrained by scattering data, the associated interaction energies denoted by $E^{2N}_{\rm NM}$ and $E^{(c_1+c_3)}_{\rm NM}$, respectively, can be calculated accurately using quantum many-body methods \cite{Hebeler:2009iv,Drischler:2020hwi,Tews:2024owl}. Noting that the symmetry energy is the difference between the energy per particle of neutron and nuclear matter at saturation density, we define 
\beq
\delta S_0 = S_0 - (\tilde E_{\rm NM}(\nsat) - E_{\rm SM}(\nsat))\,,
\eeq 
where 
\beq
\tilde E_{\rm NM}(n_B) = E^{\rm kin}_{\rm NM}(n_B) + E^{2N}_{\rm NM}(n_B)+  E^{(c_1+c_3)}_{\rm NM}(n_B)\,,
\eeq
is the neutron matter energy when contributions from $D_2$ and $F_2$ are neglected and $E_{\rm SM}(\nsat)\simeq -16$ MeV.

\begin{figure}[h]
    \centering
    \includegraphics[width=1\linewidth]{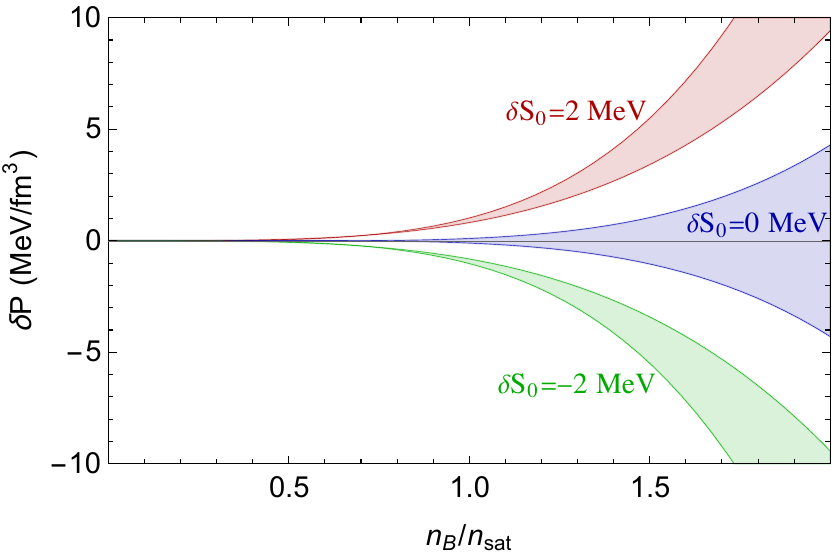}
    \caption{$D_2+F_2$ contribution to the pressure of neutron matter.}
    \label{fig:DeltaP}
\end{figure}

We correlate $D_2$ and $F_2$ by requiring that $E^D_{\rm NM}+E^F_{\rm NM}=\delta S_0$ at $n_B=\nsat$. Approximately, we find that $(100 F_2 + 9 D_2) \approx -(\delta S_0/{\rm MeV})/f_\pi^4$. 
In Fig.~\ref{fig:DeltaP} we show $\delta P$, the net contribution from the $D_2$ and $F_2$ operators to the pressure of neutron matter. The bands are obtained by varying $D_2$ in the range $\pm 1/(5 f_\pi^4)$ and $F_2$ is determined by specifying $\delta S_0$. For $\delta S_0 \simeq \pm 2$ MeV, $\delta P \simeq  \pm 1$ MeV/fm$^3$ at saturation density and the uncertainty associated with the parameter $L=3 P(\nsat)/\nsat$ is $\simeq \pm 19$ MeV. The mild discrepancy between EFT prediction for $L$ and those deduced from the neutron-skin measurement of lead using the recent parity-violating electron scattering experiment (PREX) \cite{PREX:2021umo,Reed:2021nqk} would be alleviated if this additional uncertainty is included.     

The rapid change of $\delta P$ in the density interval $\nsat -2 \nsat$ has implications for neutron stars and will be discussed in a separate paper \cite{Dekens:2024}. From  Fig.~\ref{fig:DeltaP} we deduce that earlier estimates of the uncertainty associated with Chiral EFT predictions for the pressure of neutron star matter for $n_B \simeq \nsat-2\nsat$ and its implied bounds on neutron star radii and maximum mass will need to be revised. \looseness-1

As previously noted in Ref.~\cite{Fore:2023gwv}, the couplings of pions to two nucleons are also  relevant to the analysis of pionic atoms and could explain the missing repulsion observed in phenomenological approaches~\cite{Batty:1997zp}. The pion-nucleus optical potential used to describe pionic atoms also constrains $D_2$ and $E_2$. Using the analysis in \cite{Fore:2023gwv} and several optical potentials extracted by different fitting procedures in Ref.\cite{Friedman:2002ix}, we find that $-15 ~{\rm fm}^4 <D_2 + E_2 < -1~{\rm fm}^4$. For values in this range, the s-wave interaction of the pion inside the nucleus is adequately repulsive to explain observed binding energies of pionic atoms. 

\section{Conclusions}
\label{sec:conclusions}
The contribution of the new 3NFs discussed in sections \ref{sec:matter} and \ref{sec:astro} are large enough to warrant their inclusion in all aspects of nuclear structure studies and calculations of the dense matter EOS. Using RG invariance to estimate the size of the associated LECs, we find that at saturation density, they contribute a few MeV to the energy per particle of nuclear matter and their contribution to neutron matter can be as large as $10$ MeV per particle--suggesting that earlier studies have underestimated the EFT truncation errors.  
To obtain better estimates, $D_2$ and $F_2$ should be included in fits to experimental data, along with the currently considered LECs $c_D$ and $c_E$ \cite{Lynn:2017fxg,Gazit:2008ma,Navratil:2007we,Ekstrom:2015rta,Epelbaum:2002vt,Piarulli:2017dwd,LENPIC:2018ewt,Tews:2024owl}.
We hope our study motivates the inclusion of these new 3NFs in ab initio calculations of light nuclei and look forward to stringent constraints on the associated LECs.      

$D_2$ and $F_2$ contributions are significant in neutron matter and highlight the role of using a consistent renormalization procedure in Chiral EFT. The expectation that Chiral EFT can provide a good description of neutron matter solely in terms of  LECs extracted from nucleon-nucleon and pion-nucleon scattering data would require $D_2$ and $F_2$ to be significantly smaller than suggested by the RG analysis. We combined HF calculations of $D_2$ and $F_2$ contributions to the energy of neutron matter with earlier Chiral EFT predictions to correlate and restrict these LECs with empirical information about the nuclear symmetry energy. Even for values in this restricted range, our study highlights the need to revisit EFT predictions for the density dependence of the nuclear symmetry energy and its implications for the EOS of neutron-rich matter and neutron stars, neutron skins of nuclei, and the interpretation of the data from pionic atoms. \looseness-1

\section{Acknowledgements}
Our study was supported by the U.S. Dept. of Energy under Grant No. DE-FG02-00ER41132. We thank Emmanuale Mereghetti, Bira van Kolck, and Martin Savage for useful discussions. We also thank Achim Schwenk and Epelbaum Evgeny for comments on the manuscript and their suggestions. 
\bibliographystyle{apsrev4-2}
\bibliography{bibliography}

\begin{thebibliography}{50}%
\makeatletter
\providecommand \@ifxundefined [1]{%
 \@ifx{#1\undefined}
}%
\providecommand \@ifnum [1]{%
 \ifnum #1\expandafter \@firstoftwo
 \else \expandafter \@secondoftwo
 \fi
}%
\providecommand \@ifx [1]{%
 \ifx #1\expandafter \@firstoftwo
 \else \expandafter \@secondoftwo
 \fi
}%
\providecommand \natexlab [1]{#1}%
\providecommand \enquote  [1]{``#1''}%
\providecommand \bibnamefont  [1]{#1}%
\providecommand \bibfnamefont [1]{#1}%
\providecommand \citenamefont [1]{#1}%
\providecommand \href@noop [0]{\@secondoftwo}%
\providecommand \href [0]{\begingroup \@sanitize@url \@href}%
\providecommand \@href[1]{\@@startlink{#1}\@@href}%
\providecommand \@@href[1]{\endgroup#1\@@endlink}%
\providecommand \@sanitize@url [0]{\catcode `\\12\catcode `\$12\catcode `\&12\catcode `\#12\catcode `\^12\catcode `\_12\catcode `\%12\relax}%
\providecommand \@@startlink[1]{}%
\providecommand \@@endlink[0]{}%
\providecommand \url  [0]{\begingroup\@sanitize@url \@url }%
\providecommand \@url [1]{\endgroup\@href {#1}{\urlprefix }}%
\providecommand \urlprefix  [0]{URL }%
\providecommand \Eprint [0]{\href }%
\providecommand \doibase [0]{https://doi.org/}%
\providecommand \selectlanguage [0]{\@gobble}%
\providecommand \bibinfo  [0]{\@secondoftwo}%
\providecommand \bibfield  [0]{\@secondoftwo}%
\providecommand \translation [1]{[#1]}%
\providecommand \BibitemOpen [0]{}%
\providecommand \bibitemStop [0]{}%
\providecommand \bibitemNoStop [0]{.\EOS\space}%
\providecommand \EOS [0]{\spacefactor3000\relax}%
\providecommand \BibitemShut  [1]{\csname bibitem#1\endcsname}%
\let\auto@bib@innerbib\@empty
\bibitem [{\citenamefont {Epelbaum}\ \emph {et~al.}(2009)\citenamefont {Epelbaum}, \citenamefont {Hammer},\ and\ \citenamefont {Meissner}}]{Epelbaum:2008ga}%
  \BibitemOpen
  \bibfield  {author} {\bibinfo {author} {\bibfnamefont {E.}~\bibnamefont {Epelbaum}}, \bibinfo {author} {\bibfnamefont {H.-W.}\ \bibnamefont {Hammer}},\ and\ \bibinfo {author} {\bibfnamefont {U.-G.}\ \bibnamefont {Meissner}},\ }\href {https://doi.org/10.1103/RevModPhys.81.1773} {\bibfield  {journal} {\bibinfo  {journal} {Rev. Mod. Phys.}\ }\textbf {\bibinfo {volume} {81}},\ \bibinfo {pages} {1773} (\bibinfo {year} {2009})},\ \Eprint {https://arxiv.org/abs/0811.1338} {arXiv:0811.1338 [nucl-th]} \BibitemShut {NoStop}%
\bibitem [{\citenamefont {Hammer}\ \emph {et~al.}(2020)\citenamefont {Hammer}, \citenamefont {K\"onig},\ and\ \citenamefont {van Kolck}}]{Hammer:2019poc}%
  \BibitemOpen
  \bibfield  {author} {\bibinfo {author} {\bibfnamefont {H.~W.}\ \bibnamefont {Hammer}}, \bibinfo {author} {\bibfnamefont {S.}~\bibnamefont {K\"onig}},\ and\ \bibinfo {author} {\bibfnamefont {U.}~\bibnamefont {van Kolck}},\ }\href {https://doi.org/10.1103/RevModPhys.92.025004} {\bibfield  {journal} {\bibinfo  {journal} {Rev. Mod. Phys.}\ }\textbf {\bibinfo {volume} {92}},\ \bibinfo {pages} {025004} (\bibinfo {year} {2020})},\ \Eprint {https://arxiv.org/abs/1906.12122} {arXiv:1906.12122 [nucl-th]} \BibitemShut {NoStop}%
\bibitem [{\citenamefont {Weinberg}(1990)}]{Weinberg:1990rz}%
  \BibitemOpen
  \bibfield  {author} {\bibinfo {author} {\bibfnamefont {S.}~\bibnamefont {Weinberg}},\ }\href {https://doi.org/10.1016/0370-2693(90)90938-3} {\bibfield  {journal} {\bibinfo  {journal} {Phys. Lett. B}\ }\textbf {\bibinfo {volume} {251}},\ \bibinfo {pages} {288} (\bibinfo {year} {1990})}\BibitemShut {NoStop}%
\bibitem [{\citenamefont {Weinberg}(1991)}]{Weinberg:1991um}%
  \BibitemOpen
  \bibfield  {author} {\bibinfo {author} {\bibfnamefont {S.}~\bibnamefont {Weinberg}},\ }\href {https://doi.org/10.1016/0550-3213(91)90231-L} {\bibfield  {journal} {\bibinfo  {journal} {Nucl. Phys. B}\ }\textbf {\bibinfo {volume} {363}},\ \bibinfo {pages} {3} (\bibinfo {year} {1991})}\BibitemShut {NoStop}%
\bibitem [{\citenamefont {van Kolck}(1994)}]{vanKolck:1994yi}%
  \BibitemOpen
  \bibfield  {author} {\bibinfo {author} {\bibfnamefont {U.}~\bibnamefont {van Kolck}},\ }\href {https://doi.org/10.1103/PhysRevC.49.2932} {\bibfield  {journal} {\bibinfo  {journal} {Phys. Rev. C}\ }\textbf {\bibinfo {volume} {49}},\ \bibinfo {pages} {2932} (\bibinfo {year} {1994})}\BibitemShut {NoStop}%
\bibitem [{\citenamefont {Epelbaum}\ \emph {et~al.}(2002)\citenamefont {Epelbaum}, \citenamefont {Nogga}, \citenamefont {Gloeckle}, \citenamefont {Kamada}, \citenamefont {Meissner},\ and\ \citenamefont {Witala}}]{Epelbaum:2002vt}%
  \BibitemOpen
  \bibfield  {author} {\bibinfo {author} {\bibfnamefont {E.}~\bibnamefont {Epelbaum}}, \bibinfo {author} {\bibfnamefont {A.}~\bibnamefont {Nogga}}, \bibinfo {author} {\bibfnamefont {W.}~\bibnamefont {Gloeckle}}, \bibinfo {author} {\bibfnamefont {H.}~\bibnamefont {Kamada}}, \bibinfo {author} {\bibfnamefont {U.~G.}\ \bibnamefont {Meissner}},\ and\ \bibinfo {author} {\bibfnamefont {H.}~\bibnamefont {Witala}},\ }\href {https://doi.org/10.1103/PhysRevC.66.064001} {\bibfield  {journal} {\bibinfo  {journal} {Phys. Rev. C}\ }\textbf {\bibinfo {volume} {66}},\ \bibinfo {pages} {064001} (\bibinfo {year} {2002})},\ \Eprint {https://arxiv.org/abs/nucl-th/0208023} {arXiv:nucl-th/0208023} \BibitemShut {NoStop}%
\bibitem [{\citenamefont {Fujita}\ and\ \citenamefont {Miyazawa}(1957)}]{Fujita_Miyazawa:1957}%
  \BibitemOpen
  \bibfield  {author} {\bibinfo {author} {\bibfnamefont {J.-i.}\ \bibnamefont {Fujita}}\ and\ \bibinfo {author} {\bibfnamefont {H.}~\bibnamefont {Miyazawa}},\ }\href {https://doi.org/10.1143/PTP.17.360} {\bibfield  {journal} {\bibinfo  {journal} {Progress of Theoretical Physics}\ }\textbf {\bibinfo {volume} {17}},\ \bibinfo {pages} {360} (\bibinfo {year} {1957})},\ \Eprint {https://arxiv.org/abs/https://academic.oup.com/ptp/article-pdf/17/3/360/5252121/17-3-360.pdf} {https://academic.oup.com/ptp/article-pdf/17/3/360/5252121/17-3-360.pdf} \BibitemShut {NoStop}%
\bibitem [{\citenamefont {Beane}\ \emph {et~al.}(2014)\citenamefont {Beane}, \citenamefont {Cohen}, \citenamefont {Detmold}, \citenamefont {Lin},\ and\ \citenamefont {Savage}}]{Beane:2013kca}%
  \BibitemOpen
  \bibfield  {author} {\bibinfo {author} {\bibfnamefont {S.~R.}\ \bibnamefont {Beane}}, \bibinfo {author} {\bibfnamefont {S.~D.}\ \bibnamefont {Cohen}}, \bibinfo {author} {\bibfnamefont {W.}~\bibnamefont {Detmold}}, \bibinfo {author} {\bibfnamefont {H.~W.}\ \bibnamefont {Lin}},\ and\ \bibinfo {author} {\bibfnamefont {M.~J.}\ \bibnamefont {Savage}},\ }\href {https://doi.org/10.1103/PhysRevD.89.074505} {\bibfield  {journal} {\bibinfo  {journal} {Phys. Rev. D}\ }\textbf {\bibinfo {volume} {89}},\ \bibinfo {pages} {074505} (\bibinfo {year} {2014})},\ \Eprint {https://arxiv.org/abs/1306.6939} {arXiv:1306.6939 [hep-ph]} \BibitemShut {NoStop}%
\bibitem [{\citenamefont {Kaplan}\ \emph {et~al.}(1996)\citenamefont {Kaplan}, \citenamefont {Savage},\ and\ \citenamefont {Wise}}]{Kaplan:1996xu}%
  \BibitemOpen
  \bibfield  {author} {\bibinfo {author} {\bibfnamefont {D.~B.}\ \bibnamefont {Kaplan}}, \bibinfo {author} {\bibfnamefont {M.~J.}\ \bibnamefont {Savage}},\ and\ \bibinfo {author} {\bibfnamefont {M.~B.}\ \bibnamefont {Wise}},\ }\href {https://doi.org/10.1016/0550-3213(96)00357-4} {\bibfield  {journal} {\bibinfo  {journal} {Nucl. Phys. B}\ }\textbf {\bibinfo {volume} {478}},\ \bibinfo {pages} {629} (\bibinfo {year} {1996})},\ \Eprint {https://arxiv.org/abs/nucl-th/9605002} {arXiv:nucl-th/9605002} \BibitemShut {NoStop}%
\bibitem [{\citenamefont {Mehen}\ and\ \citenamefont {Stewart}(1999)}]{Mehen:1998tp}%
  \BibitemOpen
  \bibfield  {author} {\bibinfo {author} {\bibfnamefont {T.}~\bibnamefont {Mehen}}\ and\ \bibinfo {author} {\bibfnamefont {I.~W.}\ \bibnamefont {Stewart}},\ }\href {https://doi.org/10.1103/PhysRevC.59.2365} {\bibfield  {journal} {\bibinfo  {journal} {Phys. Rev. C}\ }\textbf {\bibinfo {volume} {59}},\ \bibinfo {pages} {2365} (\bibinfo {year} {1999})},\ \Eprint {https://arxiv.org/abs/nucl-th/9809095} {arXiv:nucl-th/9809095} \BibitemShut {NoStop}%
\bibitem [{\citenamefont {Beane}\ \emph {et~al.}(2006)\citenamefont {Beane}, \citenamefont {Bedaque}, \citenamefont {Orginos},\ and\ \citenamefont {Savage}}]{Beane:2006mx}%
  \BibitemOpen
  \bibfield  {author} {\bibinfo {author} {\bibfnamefont {S.~R.}\ \bibnamefont {Beane}}, \bibinfo {author} {\bibfnamefont {P.~F.}\ \bibnamefont {Bedaque}}, \bibinfo {author} {\bibfnamefont {K.}~\bibnamefont {Orginos}},\ and\ \bibinfo {author} {\bibfnamefont {M.~J.}\ \bibnamefont {Savage}},\ }\href {https://doi.org/10.1103/PhysRevLett.97.012001} {\bibfield  {journal} {\bibinfo  {journal} {Phys. Rev. Lett.}\ }\textbf {\bibinfo {volume} {97}},\ \bibinfo {pages} {012001} (\bibinfo {year} {2006})},\ \Eprint {https://arxiv.org/abs/hep-lat/0602010} {arXiv:hep-lat/0602010} \BibitemShut {NoStop}%
\bibitem [{Note1()}]{Note1}%
  \BibitemOpen
  \bibinfo {note} {Note that Ref.~\cite {Gegelia:2004pz} under-estimates the anomalous dimension in the RHS of Eq.~\protect \eqref {eq:anom_dim_1} by a factor of four. Our direct calculation agrees with the result of Ref.\cite {Kaplan:1996xu}}\BibitemShut {NoStop}%
\bibitem [{\citenamefont {Cirigliano}\ \emph {et~al.}(2019)\citenamefont {Cirigliano}, \citenamefont {Dekens}, \citenamefont {De~Vries}, \citenamefont {Graesser}, \citenamefont {Mereghetti}, \citenamefont {Pastore}, \citenamefont {Piarulli}, \citenamefont {Van~Kolck},\ and\ \citenamefont {Wiringa}}]{Cirigliano:2019vdj}%
  \BibitemOpen
  \bibfield  {author} {\bibinfo {author} {\bibfnamefont {V.}~\bibnamefont {Cirigliano}}, \bibinfo {author} {\bibfnamefont {W.}~\bibnamefont {Dekens}}, \bibinfo {author} {\bibfnamefont {J.}~\bibnamefont {De~Vries}}, \bibinfo {author} {\bibfnamefont {M.~L.}\ \bibnamefont {Graesser}}, \bibinfo {author} {\bibfnamefont {E.}~\bibnamefont {Mereghetti}}, \bibinfo {author} {\bibfnamefont {S.}~\bibnamefont {Pastore}}, \bibinfo {author} {\bibfnamefont {M.}~\bibnamefont {Piarulli}}, \bibinfo {author} {\bibfnamefont {U.}~\bibnamefont {Van~Kolck}},\ and\ \bibinfo {author} {\bibfnamefont {R.~B.}\ \bibnamefont {Wiringa}},\ }\href {https://doi.org/10.1103/PhysRevC.100.055504} {\bibfield  {journal} {\bibinfo  {journal} {Phys. Rev. C}\ }\textbf {\bibinfo {volume} {100}},\ \bibinfo {pages} {055504} (\bibinfo {year} {2019})},\ \Eprint {https://arxiv.org/abs/1907.11254} {arXiv:1907.11254 [nucl-th]} \BibitemShut {NoStop}%
\bibitem [{\citenamefont {Beane}\ and\ \citenamefont {Savage}(2003)}]{Beane:2002xf}%
  \BibitemOpen
  \bibfield  {author} {\bibinfo {author} {\bibfnamefont {S.~R.}\ \bibnamefont {Beane}}\ and\ \bibinfo {author} {\bibfnamefont {M.~J.}\ \bibnamefont {Savage}},\ }\href {https://doi.org/10.1016/S0375-9474(02)01586-5} {\bibfield  {journal} {\bibinfo  {journal} {Nucl. Phys. A}\ }\textbf {\bibinfo {volume} {717}},\ \bibinfo {pages} {91} (\bibinfo {year} {2003})},\ \Eprint {https://arxiv.org/abs/nucl-th/0208021} {arXiv:nucl-th/0208021} \BibitemShut {NoStop}%
\bibitem [{\citenamefont {Kaplan}\ \emph {et~al.}(1998{\natexlab{a}})\citenamefont {Kaplan}, \citenamefont {Savage},\ and\ \citenamefont {Wise}}]{kaplan:1998we}%
  \BibitemOpen
  \bibfield  {author} {\bibinfo {author} {\bibfnamefont {D.~B.}\ \bibnamefont {Kaplan}}, \bibinfo {author} {\bibfnamefont {M.~J.}\ \bibnamefont {Savage}},\ and\ \bibinfo {author} {\bibfnamefont {M.~B.}\ \bibnamefont {Wise}},\ }\href {https://doi.org/10.1016/S0550-3213(98)00440-4} {\bibfield  {journal} {\bibinfo  {journal} {Nucl. Phys. B}\ }\textbf {\bibinfo {volume} {534}},\ \bibinfo {pages} {329} (\bibinfo {year} {1998}{\natexlab{a}})},\ \Eprint {https://arxiv.org/abs/nucl-th/9802075} {arXiv:nucl-th/9802075} \BibitemShut {NoStop}%
\bibitem [{\citenamefont {Kaplan}\ \emph {et~al.}(1998{\natexlab{b}})\citenamefont {Kaplan}, \citenamefont {Savage},\ and\ \citenamefont {Wise}}]{Kaplan:1998tg}%
  \BibitemOpen
  \bibfield  {author} {\bibinfo {author} {\bibfnamefont {D.~B.}\ \bibnamefont {Kaplan}}, \bibinfo {author} {\bibfnamefont {M.~J.}\ \bibnamefont {Savage}},\ and\ \bibinfo {author} {\bibfnamefont {M.~B.}\ \bibnamefont {Wise}},\ }\href {https://doi.org/10.1016/S0370-2693(98)00210-X} {\bibfield  {journal} {\bibinfo  {journal} {Phys. Lett. B}\ }\textbf {\bibinfo {volume} {424}},\ \bibinfo {pages} {390} (\bibinfo {year} {1998}{\natexlab{b}})},\ \Eprint {https://arxiv.org/abs/nucl-th/9801034} {arXiv:nucl-th/9801034} \BibitemShut {NoStop}%
\bibitem [{\citenamefont {Beane}\ \emph {et~al.}(2002)\citenamefont {Beane}, \citenamefont {Bedaque}, \citenamefont {Savage},\ and\ \citenamefont {van Kolck}}]{Beane:2001bc}%
  \BibitemOpen
  \bibfield  {author} {\bibinfo {author} {\bibfnamefont {S.~R.}\ \bibnamefont {Beane}}, \bibinfo {author} {\bibfnamefont {P.~F.}\ \bibnamefont {Bedaque}}, \bibinfo {author} {\bibfnamefont {M.~J.}\ \bibnamefont {Savage}},\ and\ \bibinfo {author} {\bibfnamefont {U.}~\bibnamefont {van Kolck}},\ }\href {https://doi.org/10.1016/S0375-9474(01)01324-0} {\bibfield  {journal} {\bibinfo  {journal} {Nucl. Phys. A}\ }\textbf {\bibinfo {volume} {700}},\ \bibinfo {pages} {377} (\bibinfo {year} {2002})},\ \Eprint {https://arxiv.org/abs/nucl-th/0104030} {arXiv:nucl-th/0104030} \BibitemShut {NoStop}%
\bibitem [{Note2()}]{Note2}%
  \BibitemOpen
  \bibinfo {note} {For example, numerically large and/or $\pi $-enhanced contributions have been encountered in $\pi -N$ and $\pi $-deuteron scattering \cite {Gasser:2002am,Hoferichter:2009ez,Hoferichter:2009gn,Beane:2002wk,Liebig:2010ki,Baru:2010xn,Baru:2011bw,Baru:2012iv}. However, there is no general power-counting that can capture such enhancements, without explicit calculation.}\BibitemShut {Stop}%
\bibitem [{\citenamefont {Borasoy}\ and\ \citenamefont {Griesshammer}(2001)}]{Borasoy:2001gq}%
  \BibitemOpen
  \bibfield  {author} {\bibinfo {author} {\bibfnamefont {B.}~\bibnamefont {Borasoy}}\ and\ \bibinfo {author} {\bibfnamefont {H.~W.}\ \bibnamefont {Griesshammer}},\ }\href@noop {} {\bibinfo {title} {{The S wave pion deuteron scattering length in effective field theory}}} (\bibinfo {year} {2001}),\ \Eprint {https://arxiv.org/abs/nucl-th/0105048} {arXiv:nucl-th/0105048} \BibitemShut {NoStop}%
\bibitem [{\citenamefont {Borasoy}\ and\ \citenamefont {Griesshammer}(2003)}]{Borasoy:2003gf}%
  \BibitemOpen
  \bibfield  {author} {\bibinfo {author} {\bibfnamefont {B.}~\bibnamefont {Borasoy}}\ and\ \bibinfo {author} {\bibfnamefont {H.~W.}\ \bibnamefont {Griesshammer}},\ }\href {https://doi.org/10.1142/S0218301303001156} {\bibfield  {journal} {\bibinfo  {journal} {Int. J. Mod. Phys. E}\ }\textbf {\bibinfo {volume} {12}},\ \bibinfo {pages} {65} (\bibinfo {year} {2003})}\BibitemShut {NoStop}%
\bibitem [{Note3()}]{Note3}%
  \BibitemOpen
  \bibinfo {note} {In principle, an isospin-tensor operator of the form, $ (N^T P_a N)(N^T P_b N)^\dagger \langle u_\mu \tau ^c\rangle \langle u^\mu \tau ^d\rangle (3\delta _{ac}\delta _{bd}-\delta _{ab}\delta _{cd})$ appears in the $^{1}S_0$ channel, in addition to the singlet term, $\sim \delta _{ab}\delta _{cd}$, in Eq.\ \protect \eqref {eq:E2F2}. In fact, we find that this operator has the same anomalous dimension as that of Eq.\ \protect \eqref {eq:E2F2RG}. We do not discuss this term any further here, as the contributions to the 3NF are only sensitive to the singlet operator.}\BibitemShut {Stop}%
\bibitem [{Note4()}]{Note4}%
  \BibitemOpen
  \bibinfo {note} {One might expect these contributions to become comparable to $D_2$ for Fermi momenta of ${\protect \cal O}(300)$ MeV, at which point $E_{\protect \rm kin}^2\sim m_\pi ^2$. However, explicit calculation shows that the $E_2$ terms only become comparable for significantly larger Fermi momenta.}\BibitemShut {Stop}%
\bibitem [{\citenamefont {Hebeler}\ and\ \citenamefont {Schwenk}(2010)}]{Hebeler:2009iv}%
  \BibitemOpen
  \bibfield  {author} {\bibinfo {author} {\bibfnamefont {K.}~\bibnamefont {Hebeler}}\ and\ \bibinfo {author} {\bibfnamefont {A.}~\bibnamefont {Schwenk}},\ }\href {https://doi.org/10.1103/PhysRevC.82.014314} {\bibfield  {journal} {\bibinfo  {journal} {Phys. Rev. C}\ }\textbf {\bibinfo {volume} {82}},\ \bibinfo {pages} {014314} (\bibinfo {year} {2010})},\ \Eprint {https://arxiv.org/abs/0911.0483} {arXiv:0911.0483 [nucl-th]} \BibitemShut {NoStop}%
\bibitem [{\citenamefont {Rijken}(1991)}]{Rijken:1990qs}%
  \BibitemOpen
  \bibfield  {author} {\bibinfo {author} {\bibfnamefont {T.~A.}\ \bibnamefont {Rijken}},\ }\href {https://doi.org/10.1016/0003-4916(91)90296-K} {\bibfield  {journal} {\bibinfo  {journal} {Annals Phys.}\ }\textbf {\bibinfo {volume} {208}},\ \bibinfo {pages} {253} (\bibinfo {year} {1991})}\BibitemShut {NoStop}%
\bibitem [{\citenamefont {Reinert}\ \emph {et~al.}(2018)\citenamefont {Reinert}, \citenamefont {Krebs},\ and\ \citenamefont {Epelbaum}}]{Reinert:2017usi}%
  \BibitemOpen
  \bibfield  {author} {\bibinfo {author} {\bibfnamefont {P.}~\bibnamefont {Reinert}}, \bibinfo {author} {\bibfnamefont {H.}~\bibnamefont {Krebs}},\ and\ \bibinfo {author} {\bibfnamefont {E.}~\bibnamefont {Epelbaum}},\ }\href {https://doi.org/10.1140/epja/i2018-12516-4} {\bibfield  {journal} {\bibinfo  {journal} {Eur. Phys. J. A}\ }\textbf {\bibinfo {volume} {54}},\ \bibinfo {pages} {86} (\bibinfo {year} {2018})},\ \Eprint {https://arxiv.org/abs/1711.08821} {arXiv:1711.08821 [nucl-th]} \BibitemShut {NoStop}%
\bibitem [{\citenamefont {Bogner}\ \emph {et~al.}(2005)\citenamefont {Bogner}, \citenamefont {Schwenk}, \citenamefont {Furnstahl},\ and\ \citenamefont {Nogga}}]{Bogner:2005sn}%
  \BibitemOpen
  \bibfield  {author} {\bibinfo {author} {\bibfnamefont {S.~K.}\ \bibnamefont {Bogner}}, \bibinfo {author} {\bibfnamefont {A.}~\bibnamefont {Schwenk}}, \bibinfo {author} {\bibfnamefont {R.~J.}\ \bibnamefont {Furnstahl}},\ and\ \bibinfo {author} {\bibfnamefont {A.}~\bibnamefont {Nogga}},\ }\href {https://doi.org/10.1016/j.nuclphysa.2005.08.024} {\bibfield  {journal} {\bibinfo  {journal} {Nucl. Phys. A}\ }\textbf {\bibinfo {volume} {763}},\ \bibinfo {pages} {59} (\bibinfo {year} {2005})},\ \Eprint {https://arxiv.org/abs/nucl-th/0504043} {arXiv:nucl-th/0504043} \BibitemShut {NoStop}%
\bibitem [{\citenamefont {Li}\ \emph {et~al.}(2019)\citenamefont {Li}, \citenamefont {Krastev}, \citenamefont {Wen},\ and\ \citenamefont {Zhang}}]{Li:2019xxz}%
  \BibitemOpen
  \bibfield  {author} {\bibinfo {author} {\bibfnamefont {B.-A.}\ \bibnamefont {Li}}, \bibinfo {author} {\bibfnamefont {P.~G.}\ \bibnamefont {Krastev}}, \bibinfo {author} {\bibfnamefont {D.-H.}\ \bibnamefont {Wen}},\ and\ \bibinfo {author} {\bibfnamefont {N.-B.}\ \bibnamefont {Zhang}},\ }\href {https://doi.org/10.1140/epja/i2019-12780-8} {\bibfield  {journal} {\bibinfo  {journal} {Eur. Phys. J. A}\ }\textbf {\bibinfo {volume} {55}},\ \bibinfo {pages} {117} (\bibinfo {year} {2019})},\ \Eprint {https://arxiv.org/abs/1905.13175} {arXiv:1905.13175 [nucl-th]} \BibitemShut {NoStop}%
\bibitem [{\citenamefont {Drischler}\ \emph {et~al.}(2020)\citenamefont {Drischler}, \citenamefont {Furnstahl}, \citenamefont {Melendez},\ and\ \citenamefont {Phillips}}]{Drischler:2020hwi}%
  \BibitemOpen
  \bibfield  {author} {\bibinfo {author} {\bibfnamefont {C.}~\bibnamefont {Drischler}}, \bibinfo {author} {\bibfnamefont {R.~J.}\ \bibnamefont {Furnstahl}}, \bibinfo {author} {\bibfnamefont {J.~A.}\ \bibnamefont {Melendez}},\ and\ \bibinfo {author} {\bibfnamefont {D.~R.}\ \bibnamefont {Phillips}},\ }\href {https://doi.org/10.1103/PhysRevLett.125.202702} {\bibfield  {journal} {\bibinfo  {journal} {Phys. Rev. Lett.}\ }\textbf {\bibinfo {volume} {125}},\ \bibinfo {pages} {202702} (\bibinfo {year} {2020})},\ \Eprint {https://arxiv.org/abs/2004.07232} {arXiv:2004.07232 [nucl-th]} \BibitemShut {NoStop}%
\bibitem [{\citenamefont {Tews}\ \emph {et~al.}(2024)\citenamefont {Tews}, \citenamefont {Somasundaram}, \citenamefont {Lonardoni}, \citenamefont {G\"ottling}, \citenamefont {Seutin}, \citenamefont {Carlson}, \citenamefont {Gandolfi}, \citenamefont {Hebeler},\ and\ \citenamefont {Schwenk}}]{Tews:2024owl}%
  \BibitemOpen
  \bibfield  {author} {\bibinfo {author} {\bibfnamefont {I.}~\bibnamefont {Tews}}, \bibinfo {author} {\bibfnamefont {R.}~\bibnamefont {Somasundaram}}, \bibinfo {author} {\bibfnamefont {D.}~\bibnamefont {Lonardoni}}, \bibinfo {author} {\bibfnamefont {H.}~\bibnamefont {G\"ottling}}, \bibinfo {author} {\bibfnamefont {R.}~\bibnamefont {Seutin}}, \bibinfo {author} {\bibfnamefont {J.}~\bibnamefont {Carlson}}, \bibinfo {author} {\bibfnamefont {S.}~\bibnamefont {Gandolfi}}, \bibinfo {author} {\bibfnamefont {K.}~\bibnamefont {Hebeler}},\ and\ \bibinfo {author} {\bibfnamefont {A.}~\bibnamefont {Schwenk}},\ }\href@noop {} {\bibinfo {title} {{Neutron matter from local chiral EFT interactions at large cutoffs}}} (\bibinfo {year} {2024}),\ \Eprint {https://arxiv.org/abs/2407.08979} {arXiv:2407.08979 [nucl-th]} \BibitemShut {NoStop}%
\bibitem [{\citenamefont {Adhikari}\ \emph {et~al.}(2021)\citenamefont {Adhikari} \emph {et~al.}}]{PREX:2021umo}%
  \BibitemOpen
  \bibfield  {author} {\bibinfo {author} {\bibfnamefont {D.}~\bibnamefont {Adhikari}} \emph {et~al.} (\bibinfo {collaboration} {PREX}),\ }\href {https://doi.org/10.1103/PhysRevLett.126.172502} {\bibfield  {journal} {\bibinfo  {journal} {Phys. Rev. Lett.}\ }\textbf {\bibinfo {volume} {126}},\ \bibinfo {pages} {172502} (\bibinfo {year} {2021})},\ \Eprint {https://arxiv.org/abs/2102.10767} {arXiv:2102.10767 [nucl-ex]} \BibitemShut {NoStop}%
\bibitem [{\citenamefont {Reed}\ \emph {et~al.}(2021)\citenamefont {Reed}, \citenamefont {Fattoyev}, \citenamefont {Horowitz},\ and\ \citenamefont {Piekarewicz}}]{Reed:2021nqk}%
  \BibitemOpen
  \bibfield  {author} {\bibinfo {author} {\bibfnamefont {B.~T.}\ \bibnamefont {Reed}}, \bibinfo {author} {\bibfnamefont {F.~J.}\ \bibnamefont {Fattoyev}}, \bibinfo {author} {\bibfnamefont {C.~J.}\ \bibnamefont {Horowitz}},\ and\ \bibinfo {author} {\bibfnamefont {J.}~\bibnamefont {Piekarewicz}},\ }\href {https://doi.org/10.1103/PhysRevLett.126.172503} {\bibfield  {journal} {\bibinfo  {journal} {Phys. Rev. Lett.}\ }\textbf {\bibinfo {volume} {126}},\ \bibinfo {pages} {172503} (\bibinfo {year} {2021})},\ \Eprint {https://arxiv.org/abs/2101.03193} {arXiv:2101.03193 [nucl-th]} \BibitemShut {NoStop}%
\bibitem [{\citenamefont {Dawid}\ \emph {et~al.}(2024)\citenamefont {Dawid}, \citenamefont {Dekens}, \citenamefont {Drischler}, \citenamefont {Kumamoto},\ and\ \citenamefont {Reddy}}]{Dekens:2024}%
  \BibitemOpen
  \bibfield  {author} {\bibinfo {author} {\bibfnamefont {M.}~\bibnamefont {Dawid}}, \bibinfo {author} {\bibfnamefont {W.}~\bibnamefont {Dekens}}, \bibinfo {author} {\bibfnamefont {C.}~\bibnamefont {Drischler}}, \bibinfo {author} {\bibfnamefont {M.}~\bibnamefont {Kumamoto}},\ and\ \bibinfo {author} {\bibfnamefont {S.}~\bibnamefont {Reddy}},\ }\href@noop {} {\bibfield  {journal} {\bibinfo  {journal} {in preperation}\ } (\bibinfo {year} {2024})}\BibitemShut {NoStop}%
\bibitem [{\citenamefont {Fore}\ \emph {et~al.}(2024)\citenamefont {Fore}, \citenamefont {Kaiser}, \citenamefont {Reddy},\ and\ \citenamefont {Warrington}}]{Fore:2023gwv}%
  \BibitemOpen
  \bibfield  {author} {\bibinfo {author} {\bibfnamefont {B.}~\bibnamefont {Fore}}, \bibinfo {author} {\bibfnamefont {N.}~\bibnamefont {Kaiser}}, \bibinfo {author} {\bibfnamefont {S.}~\bibnamefont {Reddy}},\ and\ \bibinfo {author} {\bibfnamefont {N.~C.}\ \bibnamefont {Warrington}},\ }\href {https://doi.org/10.1103/PhysRevC.110.025803} {\bibfield  {journal} {\bibinfo  {journal} {Phys. Rev. C}\ }\textbf {\bibinfo {volume} {110}},\ \bibinfo {pages} {025803} (\bibinfo {year} {2024})},\ \Eprint {https://arxiv.org/abs/2301.07226} {arXiv:2301.07226 [nucl-th]} \BibitemShut {NoStop}%
\bibitem [{\citenamefont {Batty}\ \emph {et~al.}(1997)\citenamefont {Batty}, \citenamefont {Friedman},\ and\ \citenamefont {Gal}}]{Batty:1997zp}%
  \BibitemOpen
  \bibfield  {author} {\bibinfo {author} {\bibfnamefont {C.~J.}\ \bibnamefont {Batty}}, \bibinfo {author} {\bibfnamefont {E.}~\bibnamefont {Friedman}},\ and\ \bibinfo {author} {\bibfnamefont {A.}~\bibnamefont {Gal}},\ }\href {https://doi.org/10.1016/S0370-1573(97)00011-2} {\bibfield  {journal} {\bibinfo  {journal} {Phys. Rept.}\ }\textbf {\bibinfo {volume} {287}},\ \bibinfo {pages} {385} (\bibinfo {year} {1997})}\BibitemShut {NoStop}%
\bibitem [{\citenamefont {Friedman}(2002)}]{Friedman:2002ix}%
  \BibitemOpen
  \bibfield  {author} {\bibinfo {author} {\bibfnamefont {E.}~\bibnamefont {Friedman}},\ }\href {https://doi.org/10.1016/S0375-9474(02)01126-0} {\bibfield  {journal} {\bibinfo  {journal} {Nucl. Phys. A}\ }\textbf {\bibinfo {volume} {710}},\ \bibinfo {pages} {117} (\bibinfo {year} {2002})},\ \Eprint {https://arxiv.org/abs/nucl-th/0205054} {arXiv:nucl-th/0205054} \BibitemShut {NoStop}%
\bibitem [{\citenamefont {Lynn}\ \emph {et~al.}(2017)\citenamefont {Lynn}, \citenamefont {Tews}, \citenamefont {Carlson}, \citenamefont {Gandolfi}, \citenamefont {Gezerlis}, \citenamefont {Schmidt},\ and\ \citenamefont {Schwenk}}]{Lynn:2017fxg}%
  \BibitemOpen
  \bibfield  {author} {\bibinfo {author} {\bibfnamefont {J.~E.}\ \bibnamefont {Lynn}}, \bibinfo {author} {\bibfnamefont {I.}~\bibnamefont {Tews}}, \bibinfo {author} {\bibfnamefont {J.}~\bibnamefont {Carlson}}, \bibinfo {author} {\bibfnamefont {S.}~\bibnamefont {Gandolfi}}, \bibinfo {author} {\bibfnamefont {A.}~\bibnamefont {Gezerlis}}, \bibinfo {author} {\bibfnamefont {K.~E.}\ \bibnamefont {Schmidt}},\ and\ \bibinfo {author} {\bibfnamefont {A.}~\bibnamefont {Schwenk}},\ }\href {https://doi.org/10.1103/PhysRevC.96.054007} {\bibfield  {journal} {\bibinfo  {journal} {Phys. Rev. C}\ }\textbf {\bibinfo {volume} {96}},\ \bibinfo {pages} {054007} (\bibinfo {year} {2017})},\ \Eprint {https://arxiv.org/abs/1706.07668} {arXiv:1706.07668 [nucl-th]} \BibitemShut {NoStop}%
\bibitem [{\citenamefont {Gazit}\ \emph {et~al.}(2009)\citenamefont {Gazit}, \citenamefont {Quaglioni},\ and\ \citenamefont {Navratil}}]{Gazit:2008ma}%
  \BibitemOpen
  \bibfield  {author} {\bibinfo {author} {\bibfnamefont {D.}~\bibnamefont {Gazit}}, \bibinfo {author} {\bibfnamefont {S.}~\bibnamefont {Quaglioni}},\ and\ \bibinfo {author} {\bibfnamefont {P.}~\bibnamefont {Navratil}},\ }\href {https://doi.org/10.1103/PhysRevLett.103.102502} {\bibfield  {journal} {\bibinfo  {journal} {Phys. Rev. Lett.}\ }\textbf {\bibinfo {volume} {103}},\ \bibinfo {pages} {102502} (\bibinfo {year} {2009})},\ \bibinfo {note} {[Erratum: Phys.Rev.Lett. 122, 029901 (2019)]},\ \Eprint {https://arxiv.org/abs/0812.4444} {arXiv:0812.4444 [nucl-th]} \BibitemShut {NoStop}%
\bibitem [{\citenamefont {Navratil}\ \emph {et~al.}(2007)\citenamefont {Navratil}, \citenamefont {Gueorguiev}, \citenamefont {Vary}, \citenamefont {Ormand},\ and\ \citenamefont {Nogga}}]{Navratil:2007we}%
  \BibitemOpen
  \bibfield  {author} {\bibinfo {author} {\bibfnamefont {P.}~\bibnamefont {Navratil}}, \bibinfo {author} {\bibfnamefont {V.~G.}\ \bibnamefont {Gueorguiev}}, \bibinfo {author} {\bibfnamefont {J.~P.}\ \bibnamefont {Vary}}, \bibinfo {author} {\bibfnamefont {W.~E.}\ \bibnamefont {Ormand}},\ and\ \bibinfo {author} {\bibfnamefont {A.}~\bibnamefont {Nogga}},\ }\href {https://doi.org/10.1103/PhysRevLett.99.042501} {\bibfield  {journal} {\bibinfo  {journal} {Phys. Rev. Lett.}\ }\textbf {\bibinfo {volume} {99}},\ \bibinfo {pages} {042501} (\bibinfo {year} {2007})},\ \Eprint {https://arxiv.org/abs/nucl-th/0701038} {arXiv:nucl-th/0701038} \BibitemShut {NoStop}%
\bibitem [{\citenamefont {Ekstr\"om}\ \emph {et~al.}(2015)\citenamefont {Ekstr\"om}, \citenamefont {Jansen}, \citenamefont {Wendt}, \citenamefont {Hagen}, \citenamefont {Papenbrock}, \citenamefont {Carlsson}, \citenamefont {Forss\'en}, \citenamefont {Hjorth-Jensen}, \citenamefont {Navr\'atil},\ and\ \citenamefont {Nazarewicz}}]{Ekstrom:2015rta}%
  \BibitemOpen
  \bibfield  {author} {\bibinfo {author} {\bibfnamefont {A.}~\bibnamefont {Ekstr\"om}}, \bibinfo {author} {\bibfnamefont {G.~R.}\ \bibnamefont {Jansen}}, \bibinfo {author} {\bibfnamefont {K.~A.}\ \bibnamefont {Wendt}}, \bibinfo {author} {\bibfnamefont {G.}~\bibnamefont {Hagen}}, \bibinfo {author} {\bibfnamefont {T.}~\bibnamefont {Papenbrock}}, \bibinfo {author} {\bibfnamefont {B.~D.}\ \bibnamefont {Carlsson}}, \bibinfo {author} {\bibfnamefont {C.}~\bibnamefont {Forss\'en}}, \bibinfo {author} {\bibfnamefont {M.}~\bibnamefont {Hjorth-Jensen}}, \bibinfo {author} {\bibfnamefont {P.}~\bibnamefont {Navr\'atil}},\ and\ \bibinfo {author} {\bibfnamefont {W.}~\bibnamefont {Nazarewicz}},\ }\href {https://doi.org/10.1103/PhysRevC.109.059901} {\bibfield  {journal} {\bibinfo  {journal} {Phys. Rev. C}\ }\textbf {\bibinfo {volume} {91}},\ \bibinfo {pages} {051301} (\bibinfo {year} {2015})},\ \bibinfo {note} {[Erratum: Phys.Rev.C 109, 059901 (2024)]},\ \Eprint {https://arxiv.org/abs/1502.04682} {arXiv:1502.04682 [nucl-th]}
  \BibitemShut {NoStop}%
\bibitem [{\citenamefont {Piarulli}\ \emph {et~al.}(2018)\citenamefont {Piarulli} \emph {et~al.}}]{Piarulli:2017dwd}%
  \BibitemOpen
  \bibfield  {author} {\bibinfo {author} {\bibfnamefont {M.}~\bibnamefont {Piarulli}} \emph {et~al.},\ }\href {https://doi.org/10.1103/PhysRevLett.120.052503} {\bibfield  {journal} {\bibinfo  {journal} {Phys. Rev. Lett.}\ }\textbf {\bibinfo {volume} {120}},\ \bibinfo {pages} {052503} (\bibinfo {year} {2018})},\ \Eprint {https://arxiv.org/abs/1707.02883} {arXiv:1707.02883 [nucl-th]} \BibitemShut {NoStop}%
\bibitem [{\citenamefont {Epelbaum}\ \emph {et~al.}(2019)\citenamefont {Epelbaum} \emph {et~al.}}]{LENPIC:2018ewt}%
  \BibitemOpen
  \bibfield  {author} {\bibinfo {author} {\bibfnamefont {E.}~\bibnamefont {Epelbaum}} \emph {et~al.} (\bibinfo {collaboration} {LENPIC}),\ }\href {https://doi.org/10.1103/PhysRevC.99.024313} {\bibfield  {journal} {\bibinfo  {journal} {Phys. Rev. C}\ }\textbf {\bibinfo {volume} {99}},\ \bibinfo {pages} {024313} (\bibinfo {year} {2019})},\ \Eprint {https://arxiv.org/abs/1807.02848} {arXiv:1807.02848 [nucl-th]} \BibitemShut {NoStop}%
\bibitem [{\citenamefont {Gegelia}\ and\ \citenamefont {Scherer}(2006)}]{Gegelia:2004pz}%
  \BibitemOpen
  \bibfield  {author} {\bibinfo {author} {\bibfnamefont {J.}~\bibnamefont {Gegelia}}\ and\ \bibinfo {author} {\bibfnamefont {S.}~\bibnamefont {Scherer}},\ }\href {https://doi.org/10.1142/S0217751X06025237} {\bibfield  {journal} {\bibinfo  {journal} {Int. J. Mod. Phys. A}\ }\textbf {\bibinfo {volume} {21}},\ \bibinfo {pages} {1079} (\bibinfo {year} {2006})},\ \Eprint {https://arxiv.org/abs/nucl-th/0403052} {arXiv:nucl-th/0403052} \BibitemShut {NoStop}%
\bibitem [{\citenamefont {Gasser}\ \emph {et~al.}(2002)\citenamefont {Gasser}, \citenamefont {Ivanov}, \citenamefont {Lipartia}, \citenamefont {Mojzis},\ and\ \citenamefont {Rusetsky}}]{Gasser:2002am}%
  \BibitemOpen
  \bibfield  {author} {\bibinfo {author} {\bibfnamefont {J.}~\bibnamefont {Gasser}}, \bibinfo {author} {\bibfnamefont {M.~A.}\ \bibnamefont {Ivanov}}, \bibinfo {author} {\bibfnamefont {E.}~\bibnamefont {Lipartia}}, \bibinfo {author} {\bibfnamefont {M.}~\bibnamefont {Mojzis}},\ and\ \bibinfo {author} {\bibfnamefont {A.}~\bibnamefont {Rusetsky}},\ }\href {https://doi.org/10.1007/s10052-002-1013-z} {\bibfield  {journal} {\bibinfo  {journal} {Eur. Phys. J. C}\ }\textbf {\bibinfo {volume} {26}},\ \bibinfo {pages} {13} (\bibinfo {year} {2002})},\ \Eprint {https://arxiv.org/abs/hep-ph/0206068} {arXiv:hep-ph/0206068} \BibitemShut {NoStop}%
\bibitem [{\citenamefont {Hoferichter}\ \emph {et~al.}(2009)\citenamefont {Hoferichter}, \citenamefont {Kubis},\ and\ \citenamefont {Mei{\ss}ner}}]{Hoferichter:2009ez}%
  \BibitemOpen
  \bibfield  {author} {\bibinfo {author} {\bibfnamefont {M.}~\bibnamefont {Hoferichter}}, \bibinfo {author} {\bibfnamefont {B.}~\bibnamefont {Kubis}},\ and\ \bibinfo {author} {\bibfnamefont {U.-G.}\ \bibnamefont {Mei{\ss}ner}},\ }\href {https://doi.org/10.1016/j.physletb.2009.05.068} {\bibfield  {journal} {\bibinfo  {journal} {Phys. Lett. B}\ }\textbf {\bibinfo {volume} {678}},\ \bibinfo {pages} {65} (\bibinfo {year} {2009})},\ \Eprint {https://arxiv.org/abs/0903.3890} {arXiv:0903.3890 [hep-ph]} \BibitemShut {NoStop}%
\bibitem [{\citenamefont {Hoferichter}\ \emph {et~al.}(2010)\citenamefont {Hoferichter}, \citenamefont {Kubis},\ and\ \citenamefont {Mei{\ss}ner}}]{Hoferichter:2009gn}%
  \BibitemOpen
  \bibfield  {author} {\bibinfo {author} {\bibfnamefont {M.}~\bibnamefont {Hoferichter}}, \bibinfo {author} {\bibfnamefont {B.}~\bibnamefont {Kubis}},\ and\ \bibinfo {author} {\bibfnamefont {U.-G.}\ \bibnamefont {Mei{\ss}ner}},\ }\href {https://doi.org/10.1016/j.nuclphysa.2009.11.012} {\bibfield  {journal} {\bibinfo  {journal} {Nucl. Phys. A}\ }\textbf {\bibinfo {volume} {833}},\ \bibinfo {pages} {18} (\bibinfo {year} {2010})},\ \Eprint {https://arxiv.org/abs/0909.4390} {arXiv:0909.4390 [hep-ph]} \BibitemShut {NoStop}%
\bibitem [{\citenamefont {Beane}\ \emph {et~al.}(2003)\citenamefont {Beane}, \citenamefont {Bernard}, \citenamefont {Epelbaum}, \citenamefont {Mei{\ss}ner},\ and\ \citenamefont {Phillips}}]{Beane:2002wk}%
  \BibitemOpen
  \bibfield  {author} {\bibinfo {author} {\bibfnamefont {S.~R.}\ \bibnamefont {Beane}}, \bibinfo {author} {\bibfnamefont {V.}~\bibnamefont {Bernard}}, \bibinfo {author} {\bibfnamefont {E.}~\bibnamefont {Epelbaum}}, \bibinfo {author} {\bibfnamefont {U.-G.}\ \bibnamefont {Mei{\ss}ner}},\ and\ \bibinfo {author} {\bibfnamefont {D.~R.}\ \bibnamefont {Phillips}},\ }\href {https://doi.org/10.1016/S0375-9474(03)01008-X} {\bibfield  {journal} {\bibinfo  {journal} {Nucl. Phys. A}\ }\textbf {\bibinfo {volume} {720}},\ \bibinfo {pages} {399} (\bibinfo {year} {2003})},\ \Eprint {https://arxiv.org/abs/hep-ph/0206219} {arXiv:hep-ph/0206219} \BibitemShut {NoStop}%
\bibitem [{\citenamefont {Liebig}\ \emph {et~al.}(2011)\citenamefont {Liebig}, \citenamefont {Baru}, \citenamefont {Ballout}, \citenamefont {Hanhart},\ and\ \citenamefont {Nogga}}]{Liebig:2010ki}%
  \BibitemOpen
  \bibfield  {author} {\bibinfo {author} {\bibfnamefont {S.}~\bibnamefont {Liebig}}, \bibinfo {author} {\bibfnamefont {V.}~\bibnamefont {Baru}}, \bibinfo {author} {\bibfnamefont {F.}~\bibnamefont {Ballout}}, \bibinfo {author} {\bibfnamefont {C.}~\bibnamefont {Hanhart}},\ and\ \bibinfo {author} {\bibfnamefont {A.}~\bibnamefont {Nogga}},\ }\href {https://doi.org/10.1140/epja/i2011-11069-4} {\bibfield  {journal} {\bibinfo  {journal} {Eur. Phys. J. A}\ }\textbf {\bibinfo {volume} {47}},\ \bibinfo {pages} {69} (\bibinfo {year} {2011})},\ \Eprint {https://arxiv.org/abs/1003.3826} {arXiv:1003.3826 [nucl-th]} \BibitemShut {NoStop}%
\bibitem [{\citenamefont {Baru}\ \emph {et~al.}(2011{\natexlab{a}})\citenamefont {Baru}, \citenamefont {Hanhart}, \citenamefont {Hoferichter}, \citenamefont {Kubis}, \citenamefont {Nogga},\ and\ \citenamefont {Phillips}}]{Baru:2010xn}%
  \BibitemOpen
  \bibfield  {author} {\bibinfo {author} {\bibfnamefont {V.}~\bibnamefont {Baru}}, \bibinfo {author} {\bibfnamefont {C.}~\bibnamefont {Hanhart}}, \bibinfo {author} {\bibfnamefont {M.}~\bibnamefont {Hoferichter}}, \bibinfo {author} {\bibfnamefont {B.}~\bibnamefont {Kubis}}, \bibinfo {author} {\bibfnamefont {A.}~\bibnamefont {Nogga}},\ and\ \bibinfo {author} {\bibfnamefont {D.~R.}\ \bibnamefont {Phillips}},\ }\href {https://doi.org/10.1016/j.physletb.2010.10.028} {\bibfield  {journal} {\bibinfo  {journal} {Phys. Lett. B}\ }\textbf {\bibinfo {volume} {694}},\ \bibinfo {pages} {473} (\bibinfo {year} {2011}{\natexlab{a}})},\ \Eprint {https://arxiv.org/abs/1003.4444} {arXiv:1003.4444 [nucl-th]} \BibitemShut {NoStop}%
\bibitem [{\citenamefont {Baru}\ \emph {et~al.}(2011{\natexlab{b}})\citenamefont {Baru}, \citenamefont {Hanhart}, \citenamefont {Hoferichter}, \citenamefont {Kubis}, \citenamefont {Nogga},\ and\ \citenamefont {Phillips}}]{Baru:2011bw}%
  \BibitemOpen
  \bibfield  {author} {\bibinfo {author} {\bibfnamefont {V.}~\bibnamefont {Baru}}, \bibinfo {author} {\bibfnamefont {C.}~\bibnamefont {Hanhart}}, \bibinfo {author} {\bibfnamefont {M.}~\bibnamefont {Hoferichter}}, \bibinfo {author} {\bibfnamefont {B.}~\bibnamefont {Kubis}}, \bibinfo {author} {\bibfnamefont {A.}~\bibnamefont {Nogga}},\ and\ \bibinfo {author} {\bibfnamefont {D.~R.}\ \bibnamefont {Phillips}},\ }\href {https://doi.org/10.1016/j.nuclphysa.2011.09.015} {\bibfield  {journal} {\bibinfo  {journal} {Nucl. Phys. A}\ }\textbf {\bibinfo {volume} {872}},\ \bibinfo {pages} {69} (\bibinfo {year} {2011}{\natexlab{b}})},\ \Eprint {https://arxiv.org/abs/1107.5509} {arXiv:1107.5509 [nucl-th]} \BibitemShut {NoStop}%
\bibitem [{\citenamefont {Baru}\ \emph {et~al.}(2012)\citenamefont {Baru}, \citenamefont {Epelbaum}, \citenamefont {Hanhart}, \citenamefont {Hoferichter}, \citenamefont {Kudryavtsev},\ and\ \citenamefont {Phillips}}]{Baru:2012iv}%
  \BibitemOpen
  \bibfield  {author} {\bibinfo {author} {\bibfnamefont {V.}~\bibnamefont {Baru}}, \bibinfo {author} {\bibfnamefont {E.}~\bibnamefont {Epelbaum}}, \bibinfo {author} {\bibfnamefont {C.}~\bibnamefont {Hanhart}}, \bibinfo {author} {\bibfnamefont {M.}~\bibnamefont {Hoferichter}}, \bibinfo {author} {\bibfnamefont {A.~E.}\ \bibnamefont {Kudryavtsev}},\ and\ \bibinfo {author} {\bibfnamefont {D.~R.}\ \bibnamefont {Phillips}},\ }\href {https://doi.org/10.1140/epja/i2012-12069-6} {\bibfield  {journal} {\bibinfo  {journal} {Eur. Phys. J. A}\ }\textbf {\bibinfo {volume} {48}},\ \bibinfo {pages} {69} (\bibinfo {year} {2012})},\ \Eprint {https://arxiv.org/abs/1202.0208} {arXiv:1202.0208 [nucl-th]} \BibitemShut {NoStop}%
\end{thebibliography}%

\end{document}